\documentclass[aps,prl,reprint]{revtex4-2}
\usepackage{graphicx, amsmath, amssymb, hyperref}
\usepackage[caption=false]{subfig}

\begin{document}


\newcommand{\comb}[2]{{\begin{pmatrix} #1 \\ #2 \end{pmatrix}}}
\newcommand{\braket}[2]{{\left\langle #1 \middle| #2 \right\rangle}}
\newcommand{\bra}[1]{{\left\langle #1 \right|}}
\newcommand{\ket}[1]{{\left| #1 \right\rangle}}
\newcommand{\ketbra}[2]{{\left| #1 \middle\rangle \middle \langle #2 \right|}}

\newcommand{\fref}[1]{Fig.~\ref{#1}}
\newcommand{\Fref}[1]{Figure~\ref{#1}}
\newcommand{\sref}[1]{Sec.~\ref{#1}}
\newcommand{\tref}[1]{Table~\ref{#1}}


\title{Faster Computation with the Generalized Laplacian Quantum Walk}

\author{Jonas Duda}
	\email{jonasduda@creighton.edu}
	\affiliation{Department of Physics, Creighton University, 2500 California Plaza, Omaha, NE 68178}

\author{Thomas G. Wong}
	\email{thomaswong@creighton.edu}
	\affiliation{Department of Physics, Creighton University, 2500 California Plaza, Omaha, NE 68178}

\begin{abstract}
    Quantum walks are the quantum analogues of classical random walks or Markov chains. They are universal models of quantum computing, and they underpin a variety of quantum algorithms. We prove that a continuous-time quantum walk effected by a generalized Laplacian, which can arise in spin chains, can solve a computational problem more quickly than typical quantum walks governed by the standard Laplacian or adjacency matrix. This generalized Laplacian consists of the standard Laplacian plus a real-valued multiple of the degree matrix, and we prove that as the magnitude of the multiple of the degree matrix is increased, its corresponding quantum walk can search the complete bipartite graph with multiple marked vertices in time that approaches the optimal. This raises the potential for the generalized Laplacian quantum walk to be a useful method for developing additional faster quantum algorithms.
\end{abstract}

\maketitle


\textit{Introduction}---The field of quantum computing has arisen as a potential pathway to developing computers that outperform traditional, classical computers, with factoring serving as the prototypical example of a task that a quantum computer can solve efficiently via Shor's algorithm \cite{Shor1994}, while no known efficient classical algorithm exists. While Shor's algorithm, and many other quantum algorithms, are formulated in the quantum circuit model \cite{NielsenChuang2000}, there exist many other models of quantum computation that lend themselves to easier analysis, algorithm development, or physical implementation, such as quantum Turing machines \cite{Benioff1980,Benioff1982,Deutsch1985}, measurement-based (one-way) quantum computing \cite{Raussendorf2003}, adiabatic quantum computing \cite{Aharonov2004}, and quantum walks \cite{Childs2009,Lovett2010}. Just as one can program a classical random walk on a classical computer, and it can be compiled into logic gates, the universality of these quantum computational models means they can be converted into quantum circuits. Here, we focus on quantum walks, which are quantum analogues of classical random walks \cite{Kempe2003}. Quantum walks have been used to develop quantum algorithms for a variety of tasks, such as traversing networks \cite{Childs2003}, determining element distinctness \cite{Ambainis2004}, searching \cite{SKW2003}, finding triangles in graphs \cite{MSS2005}, and evaluating boolean formulas \cite{FGG2008}. Quantum walks have been realized in a variety of physical systems, including photons in waveguide lattices \cite{Perets2008}, photons in free space \cite{Broome2010}, cold atoms in optical lattices \cite{Karski2009}, trapped ions \cite{Zahhringer2010}, and superconducting circuits \cite{Yan2019}.

One formulation of a continuous-time quantum walk arises from the kinetic energy of a particle moving in discrete space \cite{FG1998b}. That is, in quantum mechanics \cite{Griffiths2018}, a particle of mass $m$ is described by a wave function $\psi(\mathbf{r},t)$ that evolves by Schr\"odinger's equation $i \hbar \partial_t \psi = H \psi$, where we will use units where the reduced Planck constant $\hbar = 1$, and $H$ is the Hamiltonian of the system. In general,
\[ H = -\frac{\hbar^2}{2m} \nabla^2 + V, \]
where the first term is the kinetic energy and is proportional to Laplace's operator $\nabla^2$, and the second term $V$ is the potential energy. A continuous-time quantum walk is obtained by discretizing space, which can be modeled by the $N$ vertices of a graph. Labeling the vertices $\ket{1}, \ket{2}, \dots, \ket{N}$, the state $\ket{\psi(t)}$ is a superposition over these vertices, and Laplace's operator $\nabla^2$ is replaced by the discrete Laplacian $L$, which is an $N \times N$ matrix defined as
\[ L = A - D, \]
where $A$ is the adjacency matrix of the graph ($A_{ij} = 1$ if vertices $i$ and $j$ are adjacent, and 0 otherwise), and $D$ is the degree matrix of the graph (a diagonal matrix where $D_{ii}$ is the number of neighbors of vertex $i$). Then, the Hamiltonian for a \emph{Laplacian quantum walk} is
\[ H_0 = -\gamma L + V, \]
where we have grouped together the coefficients of the discrete Laplacian into a variable $\gamma$, which represents the jumping rate of the quantum walk. The Laplacian quantum walk has been used for solving decision problems \cite{FG1998b}, spatial search \cite{CG2004}, and state transfer \cite{Alvir2016}.

Another formulation of the continuous-time quantum walk is obtained by dropping the degree matrix $D$, so the walk is effected by the adjacency matrix $A$ alone. That is, the Hamiltonian is
\[ H_1 = -\gamma A + V. \]
This \emph{adjacency quantum walk} can arise in spin networks with a single excitation \cite{Bose2009}, and it was used to demonstrate an exponential speedup over classical computing \cite{Childs2003}, as well as for investigations of spatial search \cite{CG2004}, evaluating boolean formulas \cite{FGG2008}, and state transfer \cite{Alvir2016}.

Recently, a \emph{generalized Laplacian quantum walk} was introduced that can also arise in spin networks with a single excitation \cite{Wong45}. It is governed by a generalized Laplacian $L_\alpha$ that is obtained by taking the standard Laplacian $L$ and adding a multiple $\alpha \in \mathbb{R}$ of the degree matrix $D$:
\[ L_\alpha = L + \alpha D. \]
Then, the generalized Laplacian quantum walk has Hamiltonian
\begin{equation}
    \label{eq:H-alpha}
    H_\alpha = -\gamma L_\alpha + V.
\end{equation}
This encompasses the previous two quantum walks when $\alpha = 0$ and when $\alpha = 1$, since $L_0 = L$ and $L_1 = A$. It was shown that the generalized Laplacian quantum walk searches a weighted barbell graph no faster than the adjacency quantum walk \cite{Wong45,Wong41}.

\begin{figure}
    \includegraphics{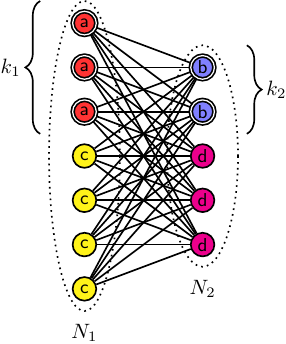}
    \caption{\label{fig:bipartite}A complete bipartite graph with $N_1 = 7$ and $N_2 = 5$ vertices in each partite set, of which $k_1 = 3$ and $k_2 = 2$ vertices are marked in the respective sets, as shown by the double circles. Identically evolving vertices are identically colored and labeled.}
\end{figure}

Here, we give a computational problem that the generalized Laplacian quantum walk does solve more quickly than the standard Laplacian and adjacency quantum walks, or any constant-$\alpha$ quantum walk. In particular, we explore search on the complete bipartite graph with multiple marked vertices, an example of which is shown in \fref{fig:bipartite}. The partite sets have $N_1$ and $N_2$ vertices, so the total number of vertices is $N = N_1 + N_2$. If the graph is regular, i.e., $N_1 = N_2$, then the degree matrix $D$ is a multiple of the identity matrix, and the quantum walk's behavior is independent of $\alpha$, up to a global phase. So, we assume that the graph is irregular, i.e., $N_1 \ne N_2$, so that the quantum walk depends on $\alpha$. In the respective partite sets, $k_1$ and $k_2$ vertices are marked by an oracle, depicted in \fref{fig:bipartite} by double circles. In a continuous-time quantum walk, the oracle manifests as a potential energy term in the Hamiltonian \cite{CG2004,Mochon2007}:
\begin{equation}
    \label{eq:oracle}
    V = -\sum_{i \in \text{marked}} \ketbra{i}{i}.
\end{equation}
The goal is to find any one of the marked vertices with as few oracle queries as possible, which in the context of a Hamiltonian oracle, means evolving by $H_\alpha$ \eqref{eq:H-alpha} for as little time as possible. Since the Hamiltonian is time-independent, the solution to Schr\"odinger's equation is
\[ \ket{\psi(t)} = e^{-iH_\alpha t} \ket{\psi(0)}, \]
and following \cite{Wong43}, for the initial state $\ket{\psi(0)}$, we use the principal eigenvector of the operator driving the quantum walk, which in this case is the generalized Laplacian $L_\alpha$.

\begin{figure}
\begin{center}
    \subfloat[$N_1 = 5000$, $N_2 = 2000$, $k_1 = 2$, and $k_2 = 3$, so $N_1 \gg N_2$.
] {
        \includegraphics{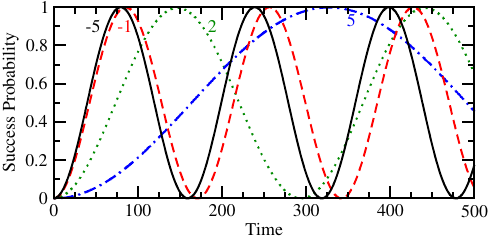}
        \label{fig:prob-time-alpha-n1big}
    }

    \subfloat[$N_1 = 2000$, $N_2 = 5000$, $k_1 = 2$, and $k_2 = 3$, so $N_1 \ll N_2$.] {
        \includegraphics{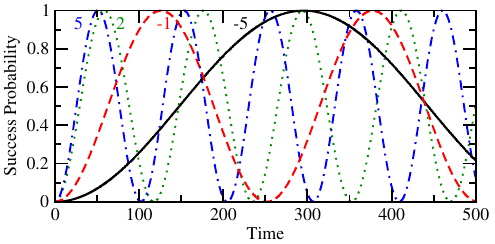}
        \label{fig:prob-time-alpha-n2big}
    }
    
    \caption{\label{fig:prob-time-alpha}Success probability versus time for search on complete bipartite graphs at $\gamma = \gamma_a $\eqref{eq:gamma-a}. The solid black curve is when $\alpha = -5$, the red dashed curve is when $\alpha = -1$, the green dotted curve is when $\alpha = 2$ and the blue dot-dashed curve is when $\alpha = 5$.}
\end{center}
\end{figure}

In \fref{fig:prob-time-alpha}, we plot the probability of finding the walker at a marked vertex, i.e., the success probability, as it evolves with time, for various values of $\alpha$, at a ``critical'' jumping rate we will derive later. In \fref{fig:prob-time-alpha-n1big}, $N_1 > N_2$, and we see that as $\alpha$ becomes more and more negative, the time for the success probability to reach 1, i.e., the runtime, gets shorter and shorter. In \fref{fig:prob-time-alpha-n2big}, $N_1 < N_2$, and now the runtime gets shorter and shorter when $\alpha$ is more and more positive. We will prove that as $\alpha \to \pm \infty$, with the sign corresponding to $\operatorname{sgn}(N_2-N_1)$, then search on the complete bipartite graph becomes optimal in both scaling and constant factor. This is faster than typical quantum walks where $\alpha = 0$ or $\alpha = 1$, and faster than any finite $\alpha$, suggesting that the generalized Laplacian quantum walk could be used to develop other faster quantum algorithms.

In the next section, we analytically prove the behavior of the algorithm for arbitrary $\alpha \ne 1$ ($\alpha = 1$ was handled in \cite{Wong19}). Then, we show that it searches in optimal time as $\alpha \to \pm \infty$, hence proving faster computation with the generalized Laplacian than other standard continuous-time quantum walks.


\textit{Analysis}---Following \cite{Wong43}, the initial state that results in an asymptotic success probability of 1 is the principal eigenvector of the operator driving the quantum walk, which in this case is the generalized Laplacian $L_\alpha$. As proved in the supplemental material, the principal eigenvector of $L_\alpha$ is
\begin{align*}
    \ket{s_{L_\alpha}} = \frac{1}{C_-} \left( \sum_{i\text{ left}} \ket{i} + \frac{2N_1}{F_-} \sum_{i\text{ right}} \ket{i} \right),
\end{align*}
where
\begin{gather*}
    B = \sqrt{(\alpha-1)^2 (N_1-N_2)^2 + 4N_1N_2}, \displaybreak[0] \\
    C_\pm = \sqrt{N_1 + \frac{4N_1^2N_2}{F_\pm^2}}, \displaybreak[0] \\
    F_\pm = B \pm (\alpha-1)(N_1-N_2).
\end{gather*}
This means each left vertex has initial amplitude $1/C_-$, and each right vertex has initial amplitude $2N_1/(C_-F_-)$, and these do not depend on the number of marked vertices nor their placement.

Due to the symmetry of the problem, the (un)marked vertices in the left set evolve identically to each other, as do the (un)marked vertices in the right set. This is depicted in \fref{fig:bipartite}, where vertices that evolve identically to each other share the same label and color. Then, the system evolves in a four-dimensional (4D) subspace with basis vectors consisting of uniform superpositions of identically-evolving vertices:
\begin{align*}
    \ket{a} &= \frac{1}{\sqrt{k_1}} \sum_{\substack{i\text{ left} \\ \text{marked}}} \ket{i}, \\
    \ket{b} &= \frac{1}{\sqrt{k_2}} \sum_{\substack{i\text{ right} \\ \text{marked}}} \ket{i}, \\
    \ket{c} &= \frac{1}{\sqrt{N_1-k_1}} \sum_{\substack{i\text{ left} \\ \text{unmarked}}} \ket{i}, \\
    \ket{d} &= \frac{1}{\sqrt{N_2-k_2}} \sum_{\substack{i\text{ right} \\ \text{unmarked}}} \ket{i}.
\end{align*}
In this 4D basis $\{ \ket{a}, \ket{b}, \ket{c}, \ket{d} \}$, the initial state is
\begin{align*}
    \ket{s_{L_\alpha}} 
        &= \frac{1}{C_-} \left( \sqrt{k_1} \ket{a} + \frac{2N_1\sqrt{k_2}}{F_-} \ket{b} + \sqrt{N_1-k_1} \ket{c} \right. \\
        &\qquad\qquad \left. + \frac{2N_1\sqrt{N_2-k_2}}{F_-} \ket{d} \right),
\end{align*}
the adjacency matrix is
\[
    A = \begin{pmatrix}
        0 & \sqrt{k_1 k_2} & 0 & \sqrt{k_1 N_{k2}} \\
        \sqrt{k_1k_2} & 0 & \sqrt{k_2 N_{k1}} & 0 \\
        0 & \sqrt{k_2 N_{k1}} & 0 & \sqrt{N_{k1} N_{k2}} \\
        \sqrt{k_1 N_{k2}} & 0 & \sqrt{N_{k1} N_{k2}} & 0 \\
    \end{pmatrix}, 
\]
where $N_{ki}=N_i-k_i$, the degree matrix is $D = \text{diag}(N_2,N_1,N_2,N_1)$, the oracle \eqref{eq:oracle} is $V = \text{diag}(-1,-1,0,0)$, and the search Hamiltonian \eqref{eq:H-alpha} is
\[H_\alpha = -\gamma \begin{pmatrix}
	   \frac{1}{\gamma} + N_2 \alpha_1 & \sqrt{k_1 k_2} & 0 & \sqrt{k_1 N_{k2}} \\
	   \sqrt{k_1 k_2} & \frac{1}{\gamma} + N_1 \alpha_1 & \sqrt{k_2 N_{k1}} & 0 \\
	   0 & \sqrt{k_2 N_{k1}} & N_2 \alpha_1 & \sqrt{N_{k1 }N_{k2}}\\
	   \sqrt{k_1 N_{k2}} & 0 & \sqrt{N_{k1} N_{k2}} & N_1 \alpha_1 \\
    \end{pmatrix},\]
where $\alpha_1 = \alpha - 1$.

As shown in the supplemental material, when the jumping rate takes a critical value of
\begin{equation}
    \label{eq:gamma-a}
    \gamma_a = \frac{2}{F_+},
\end{equation}
the asymptotic (large $N_1$ and $N_2$) eigenvectors of the search Hamiltonian $H_\alpha$ are
\begin{align*}
    &\ket{\psi_0} = \frac{1}{\sqrt{2}} ( \ket{a} + \ket{u} ), \displaybreak[0] \\
    &\ket{\psi_1} = \frac{1}{\sqrt{2}} ( \ket{a} - \ket{u} ), \displaybreak[0] \\
    &\ket{\psi_2} = \ket{b}, \displaybreak[0] \\
    &\ket{\psi_3} = \ket{v},
\end{align*}
where 
\begin{align*}
    &\ket{u} = \frac{1}{G_-} \left( \frac{F_-}{2\sqrt{N_1N_2}} \ket{c} + \ket{d} \right), \\
    &\ket{v} = \frac{1}{G_+} \left( \frac{-F_+}{2\sqrt{N_1N_2}} \ket{c} + \ket{d} \right), \\
    &G_\pm = \sqrt{1+\frac{F_\pm^2}{4 N_1 N_2}},
\end{align*}
and the corresponding eigenvalues (not necessarily in order from smallest to largest) are
\begin{align*}
    &E_0 = -\frac{J}{N_1 F_-} - \frac{\Delta E}{2}, \\
    &E_1 = -\frac{J}{N_1 F_-} + \frac{\Delta E}{2}, \\
    &E_2 = -1 - N_1(\alpha-1)\gamma_a, \\
    &E_3 = -\frac{1}{2} [N (\alpha-1) - B] \gamma_a,
\end{align*}
where
\begin{align*}
    J 
        &= -F_- (\alpha-1)^2 (N_1 - N_2) \\
        &\quad+ N_1[B + (\alpha-1)(3N_2 - N_1)]
\end{align*}
and
\[ \Delta E = 2\sqrt{\frac{k_1 (BF_--2N_1N_2)}{N_1 B F_-}}. \]

\begin{figure}
\begin{center}
    \includegraphics{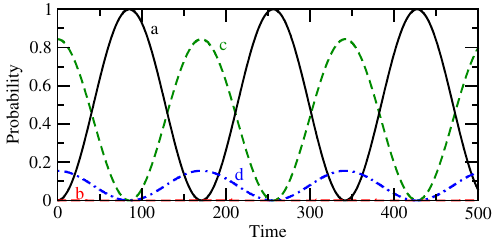}
    \caption{\label{fig:probs}The probability at each type of vertex versus time for search on the complete bipartite graph with $N_1 = 5000$, $N_2 = 2000$ vertices, of which $k_1 = 2$ and $k_2 = 3$ are marked, with $\gamma = \gamma_a$ \eqref{eq:gamma-a} $\alpha = -1$. The solid black curve is the probability in $\ket{a}$, the red dashed curve is $\ket{b}$ (which is near 0 the entire evolution), the green dotted curve is $\ket{c}$, and the blue dot-dashed curve is $\ket{d}$.}
\end{center}
\end{figure}

Now, we asymptotically express the initial state in terms of these eigenvectors. As shown in the supplemental material, $\ket{s_{L_\alpha}} \approx \ket{u}$, and so
\[ \ket{\psi(0)} = \ket{s_{L_\alpha}} \approx \ket{u} = \frac{1}{\sqrt{2}} \ket{\psi_0} - \frac{1}{\sqrt{2}} \ket{\psi_1}. \]
Then at time $t$, the state is
\begin{align*}
    \ket{\psi(t)}
        &= e^{-iHt} \ket{\psi(0)} \\
        &\approx \frac{1}{\sqrt{2}}e^{-iE_0t}\ket{\psi_0}-\frac{1}{\sqrt{2}}e^{-iE_1t}\ket{\psi_1} \displaybreak[0]  \\
        &= \frac{1}{\sqrt{2}}e^{-iE_0 t}\frac{1}{\sqrt{2}}(\ket{a}+\ket{u}) \\
            &\quad -\frac{1}{\sqrt{2}}e^{-iE_1 t}\frac{1}{\sqrt{2}}(\ket{a}-\ket{u}) \displaybreak[0]  \\
        &= \frac{1}{2}(e^{-iE_0t}-e^{-iE_1t})\ket{a}+\frac{1}{2}(e^{-iE_0t}+e^{-iE_1t})\ket{u} \displaybreak[0] \\
        &= \frac{1}{2}e^{it\frac{J}{N_1F_-}}(e^{i\Delta Et/2}-e^{-i\Delta E t/2})\ket{a} \\
            &\quad+\frac{1}{2}e^{it\frac{J}{N_1F_-}}(e^{i\Delta Et/2}+e^{-i\Delta E t/2})\ket{u} \displaybreak[0] \\
        &= e^{it\frac{J}{N_1F_-}}i\sin\left(\frac{\Delta E}{2}t\right)\ket{a} \\
            &\quad +e^{it\frac{J}{N_1F_-}}\cos\left(\frac{\Delta E}{2}t\right)\ket{u} \displaybreak[0] \\
        &= e^{it\frac{J}{N_1F_-}}i\sin\left(\frac{\Delta E}{2}t\right)\ket{a} \\
            &\quad +e^{it\frac{J}{N_1F_-}}\cos\left(\frac{\Delta E}{2}t\right)\left(\frac{F_-}{2G_-\sqrt{N_1N_2}}\right)\ket{c} \\
            &\quad+e^{it\frac{J}{N_1F_-}}\cos\left(\frac{\Delta E}{2}t\right)\frac{1}{G_-}\ket{d}.  
\end{align*}
Taking the norm-square of each amplitude, the probability at each type of vertex is
\begin{align*}
    p_a(t) &= \sin^2\left(\frac{\Delta E}{2}t\right), \displaybreak[0] \\
    p_b(t) &= 0, \displaybreak[0] \\
    p_c(t) &= \left(\frac{F_-}{2G_-\sqrt{N_1N_2}}\right)^2\cos^2\left(\frac{
    \Delta E}{2}t\right), \displaybreak[0] \\
    p_d(t) &= \frac{1}{(G_-)^2}\cos^2\left(\frac{\Delta E}{2}t\right).
\end{align*}
As a check, in \fref{fig:probs}, we plot the probability in each type of vertex over time, and it agrees with the above analytical results for large $N$. At the runtime of $t_* = \pi/\Delta E$, the probabilities reach
\[
    p_a(t_*) = 1, 
    \quad p_b(t_*) = 0, 
    \quad p_c(t_*) = 0, 
    \quad p_d(t_*) = 0,
\]
and so measuring the position of the walker at this time will result in it being found at a marked vertex in the left partite set. As a check, when $\alpha = -5, -1, 2, 5$ and with the parameters from \fref{fig:prob-time-alpha-n1big}, the runtime $t_* = \pi/\Delta E 
\approx 79.7, 85.5, 146.9, 327.0$, and with the parameters from \fref{fig:prob-time-alpha-n2big}, $\pi/\Delta E 
\approx 295.4, 125.8, 58.8, 51.2$, which are all consistent with the times at which the respective solid black, dashed red, dotted green, and dot-dashed blue success probabilities reach 1 in \fref{fig:prob-time-alpha}.

To find a marked vertex in the right partite set instead, we can leverage the symmetry of the complete bipartite graph and swap the labels from each partite set, i.e., $N_1 \leftrightarrow N_2$, $k_1 \leftrightarrow k_2$, $a \leftrightarrow b$, and $c \leftrightarrow d$ in the above results. This yields a critical jumping rate of $\gamma_b = 2/F_-$, which causes the algorithm to asymptotically evolve to $\ket{b}$, up to a phase. The exception to this is the special case where $\alpha = 1$, at which $\gamma_a = \gamma_b$, and the success probability is split between the $\ket{a}$ and $\ket{b}$ vertices rather than accumulating at $\ket{a}$ or $\ket{b}$ alone \cite{Wong19}.


\textit{Optimal Runtime}---In this section, we prove the behavior from \fref{fig:prob-time-alpha}, that when $\gamma = \gamma_a$ and $N_1 > N_2$, the runtime becomes optimal as $\alpha \to -\infty$, and when $N_1 < N_2$, it becomes optimal when $\alpha \to \infty$. Recall that the runtime occurs at time $t_* = \pi/\Delta E$. As shown in the supplemental material, the derivative of it with respect to $\alpha$ is
\[ \frac{dt_*}{d\alpha} = \frac{\pi \sqrt{2} N_1^{3/2}N_2(N_1-N_2)}{\sqrt{k_1}B^{3/2}F_-^{3/2}}. \]
Since $B$ and $F_-$ are always positive, the sign of $dt_*/d\alpha$ depends solely on $N_1 - N_2$. When $N_1 > N_2$, $dt_*/d\alpha$  is positive, and so $t_*$ is minimized when $\alpha \to -\infty$, at which $t_* \to (\pi/2)\sqrt{N_1/k_1}$. On the other hand, when $N_1 < N_2$, then $dt_*/d\alpha$ is negative, and $t_*$ is minimized when $\alpha \to \infty$, at which the runtime again becomes $(\pi/2)\sqrt{N_1/k_1}$. In both cases, as shown in the supplemental material, $\ket{s_{L_\alpha}} \to \sum_{i\text{ left}} \ket{i}/\sqrt{N_1}$, the uniform superposition over the left vertices.

Similarly, when searching for a marked vertex in the right partite set using $\gamma = \gamma_b$, $N_2 - N_1$ determines the sign of $dt_*/d\alpha$, and so $t_*$ is minimized when $\alpha \to -\infty$ when $N_2 > N_1$, and $\alpha \to \infty$ when $N_2 < N_1$, both of which result in runtimes $t_* \to (\pi/2)\sqrt{N_2/k_2}$ and initial uniform states over the right vertices, i.e., $\ket{s_{L_\alpha}} \to \sum_{i\text{ right}} \ket{i}/\sqrt{N_2}$.

Since the search problem is solved when any one marked vertex is found, the overall runtime is the minimum of the runtimes for searching the left and right marked vertices when $\alpha \to \pm \infty$, i.e.,
\[ t_* = \frac{\pi}{2} \min \left( \sqrt{\frac{N_1}{k_1}}, \sqrt{\frac{N_2}{k_2}} \right), \]
where one chooses to search the partite set with a greater density of marked vertices, i.e., the set for which $k_1/N_1$ or $k_2/N_2$ is greater. This is faster than any finite-$\alpha$ quantum walk, including the $\alpha = 1$ case from \cite{Wong19}.

For comparison, if one is classically searching for a marked vertex, then it is faster to focus on the partite set with a higher density of marked vertices, rather than on the entire complete bipartite graph. Then, the expected classical runtime is $\min(N_1/k_1, N_2/k_2)$. The fastest that a continuous-time quantum walk can search $N_i$ vertices for one of $k_i$ marked vertices occurs when the vertices are arranged in an all-to-all network, or complete graph, and this takes $(\pi/2)\sqrt{N_i/k_i}$ time \cite{Wong9}. Our generalized Laplacian quantum walk achieves this optimal runtime for the complete bipartite graph when $\alpha \to \pm \infty$, whereas quantum walks using any finite $\alpha$ are nonoptimal.


\textit{Conclusion}---We investigated searching the complete bipartite graph with the generalized Laplacian quantum walk, which has a tunable parameter $\alpha$. We derived the critical jumping rates for this graph, which can be chosen to make the system evolve from the principal eigenvector of the generalized Laplacian to to the marked vertices in either partite set. The algorithm is fastest when $\alpha \to \pm \infty$, where the sign depends on which partite set is being searched and which is larger. In this limit, an optimal algorithm is achieved, both in scaling and constant factor, outperforming the standard Laplacian and adjacency quantum walks. It may be that the generalized Laplacian quantum walk will yield additional faster quantum algorithms.


\begin{acknowledgments}
	\textit{Acknowledgments}---This material is based upon work supported in part by the National Science Foundation EPSCoR Cooperative Agreement OIA-2044049, Nebraska’s EQUATE collaboration. Any opinions, findings, and conclusions or recommendations expressed in this material are those of the author(s) and do not necessarily reflect the views of the National Science Foundation.
\end{acknowledgments}

\textit{Data availability}---The data that support the findings of this article are openly available \cite{data}.


\bibliography{refs}

\end{document}



\newcommand{\comb}[2]{{\begin{pmatrix} #1 \\ #2 \end{pmatrix}}}
\newcommand{\braket}[2]{{\left\langle #1 \middle| #2 \right\rangle}}
\newcommand{\bra}[1]{{\left\langle #1 \right|}}
\newcommand{\ket}[1]{{\left| #1 \right\rangle}}
\newcommand{\ketbra}[2]{{\left| #1 \middle\rangle \middle \langle #2 \right|}}

\newcommand{\fref}[1]{Fig.~\ref{#1}}
\newcommand{\Fref}[1]{Figure~\ref{#1}}
\newcommand{\sref}[1]{Sec.~\ref{#1}}
\newcommand{\tref}[1]{Table~\ref{#1}}


\title{Supplemental Material to \\ ``Faster Computation with the Generalized Laplacian Quantum Walk''}

\author{Jonas Duda}
	\email{jonasduda@creighton.edu}
	\affiliation{Department of Physics, Creighton University, 2500 California Plaza, Omaha, NE 68178}

\author{Thomas G. Wong}
	\email{thomaswong@creighton.edu}
	\affiliation{Department of Physics, Creighton University, 2500 California Plaza, Omaha, NE 68178}

\maketitle


\section{Exact Initial State}

In the $\{ \ket{a}, \ket{b}, \ket{c}, \ket{d} \}$ basis described in the main paper, the generalized Laplacian is
\[
    L_\alpha = \begin{pmatrix}
        (\alpha-1)N_2 & \sqrt{k_1 k_2} & 0 & \sqrt{k_1 (N_2 - k_2)} \\
        \sqrt{k_1 k_2} & (\alpha-1)N_1 & \sqrt{k_2 (N_1 - k_1)} & 0 \\
        0 & \sqrt{k_2 (N_1 - k_1)} & (\alpha-1)N_2 & \sqrt{(N_1 - k_1) (N_2 - k_2)} \\
        \sqrt{k_1 (N_2 - k_2)} & 0 & \sqrt{(N_1 - k_1) (N_2 - k_2)} & (\alpha-1)N_1 \\
    \end{pmatrix}.
\]
The eigenvalues of this are
\begin{align*}
    \lambda_0 &= \frac{1}{2} \big[ (\alpha-1)N - B \big], \\
    \lambda_1 &= (\alpha-1)N_2, \\
    \lambda_2 &= (\alpha-1)N_1,\\
    \lambda_3 &= \frac{1}{2} \big[ (\alpha-1)N + B \big],
\end{align*}
where
\[ B = \sqrt{(\alpha-1)^2(N_1-N_2)^2+4N_1N_2}. \]
Note that $\lambda_0$ is always the smallest eigenvalue, and $\lambda_3$ is always the largest eigenvalue, but the order of $\lambda_1$ and $\lambda_2$ depends on $N_1$, $N_2$, and $\alpha$. The corresponding normalized eigenvectors are
\begin{align}
    \ket{v_0} &= \frac{1}{C_+} \left( \sqrt{k_1} \ket{a} - \frac{2N_1\sqrt{k_2}}{F_+} \ket{b} + \sqrt{N_1-k_1} \ket{c} - \frac{2N_1\sqrt{N_2-k_2}}{F_+} \ket{d} \right), \nonumber \\
    \ket{v_1} &= \sqrt{\frac{N_1-k_1}{N_1}} \ket{a} - \sqrt{\frac{k_1}{N_1}} \ket{c}, \nonumber \\
    \ket{v_2} &= \sqrt{\frac{N_2-k_2}{N_2}} \ket{b} - \sqrt{\frac{k_2}{N_2}} \ket{d}, \nonumber \\
    \ket{v_3} &= \frac{1}{C_-} \left( \sqrt{k_1} \ket{a} + \frac{2N_1\sqrt{k_2}}{F_-} \ket{b} + \sqrt{N_1-k_1} \ket{c} + \frac{2N_1\sqrt{N_2-k_2}}{F_-} \ket{d} \right), \label{eq:v3}
\end{align}
where
\begin{gather*}
    C_\pm = \sqrt{N_1 + \frac{4N_1^2N_2}{F_\pm^2}}, \\
    F_\pm = B \pm (\alpha-1)(N_1-N_2).
\end{gather*}
The constants $B$, $C$, and $F$ were chosen in alphabetical order, since $A$ is already the adjacency matrix, $D$ the degree matrix, and $E$ energy. Since $\lambda_3$ is the largest eigenvalue, $\ket{v_3}$ is the principal eigenvector and is the starting state. Plugging in the definitions of $\ket{a}$, $\ket{b}$, $\ket{c}$, and $\ket{d}$, the principal eigenvector is
\begin{align}
    \ket{v_3}
        &= \frac{1}{C_-} \left( \sqrt{k_1} \frac{1}{\sqrt{k_1}} \sum_{\substack{i\text{ left} \\ \text{marked}}} \ket{i} + \frac{2N_1\sqrt{k_2}}{F_-} \frac{1}{\sqrt{k_2}} \sum_{\substack{i\text{ right} \\ \text{marked}}} \ket{i} + \sqrt{N_1-k_1} \frac{1}{\sqrt{N_1-k_1}} \sum_{\substack{i\text{ left} \\ \text{unmarked}}} \ket{i} \nonumber \right. \\
            &\qquad\qquad \left. + \frac{2N_1\sqrt{N_2-k_2}}{F_-} \frac{1}{\sqrt{N_2-k_2}} \sum_{\substack{i\text{ right} \\ \text{unmarked}}} \ket{i} \right) \nonumber \\
        &= \frac{1}{C_-} \left( \sum_{i\text{ left}} \ket{i} + \frac{2N_1}{F_-} \sum_{i\text{ right}} \ket{i} \right), \label{eq:appendix-s}
\end{align}
as reported in the main text as $\ket{s_{L_\alpha}}$.


\section{Eigenvectors of the Search Hamiltonian}

As given in the main paper, in the $\{ \ket{a}, \ket{b}, \ket{c}, \ket{d} \}$ basis, the search Hamiltonian is
\[
    H_\alpha = -\gamma \begin{pmatrix}
	   \frac{1}{\gamma} + N_2 (\alpha-1) & \sqrt{k_1 k_2} & 0 & \sqrt{k_1 (N_2 - k_2)} \\
	   \sqrt{k_1 k_2} & \frac{1}{\gamma} + N_1 (\alpha-1) & \sqrt{k_2 (N_1 - k_1)} & 0 \\
	   0 & \sqrt{k_2 (N_1 - k_1)} & N_2 (\alpha-1) & \sqrt{(N_1 - k_1) (N_2 - k_2)}\\
	   \sqrt{k_1 (N_2 - k_2)} & 0 & \sqrt{(N_1 - k_1) (N_2 - k_2)} & N_1 (\alpha-1) \\
    \end{pmatrix}.
\]
In this section, we find the eigenvectors and eigenvalues of this for large $N_1$ and $N_2$ using degenerate perturbation theory, where we first find the eigenvectors and eigenvalues of the leading order Hamiltonian $H_\alpha^{(0)}$ and then lift the degeneracy by adding the next order corrections $H_\alpha^{(1)}$. The leading-order Hamiltonian is
\[
    H_\alpha^{(0)} = -\gamma \begin{pmatrix}
	   \frac{1}{\gamma} + N_2 (\alpha-1) & 0 & 0 & 0 \\
	   0 & \frac{1}{\gamma} + N_1 (\alpha-1) & 0 & 0 \\
	   0 & 0 & N_2 (\alpha-1) & \sqrt{N_1 N_2} \\
	   0 & 0 & \sqrt{N_1 N_2} & N_1 (\alpha-1) \\
    \end{pmatrix}.
\]
The eigenvectors and eigenvalues of this are
\begin{align*}
    &\ket{a}, && -1 - N_2 (\alpha-1) \gamma, \\
    &\ket{b}, && -1 - N_1 (\alpha-1) \gamma, \\
    &\ket{u} = \frac{1}{G_-} \left( \frac{F_-}{2\sqrt{N_1N_2}} \ket{c} + \ket{d} \right), && -\frac{1}{2} [N (\alpha-1) + B] \gamma, \\
    &\ket{v} = \frac{1}{G_+} \left( \frac{-F_+}{2\sqrt{N_1N_2}} \ket{c} + \ket{d} \right), && -\frac{1}{2} [N (\alpha-1) - B] \gamma,
\end{align*}
where
\[ G_\pm = \sqrt{ 1 + \frac{F_\pm^2}{4N_1N_2} } \]
is a normalization constant. Note that $\ket{u}$ is approximately the initial state $\ket{s_{L_\alpha}} = \ket{v_3}$ \eqref{eq:v3}. That is,
\begin{align*}
    \ket{s_{L_\alpha}}
        &= \ket{v_3} \\
        &= \frac{1}{C_-} \left( \sqrt{k_1} \ket{a} + \frac{2N_1\sqrt{k_2}}{F_-} \ket{b} + \sqrt{N_1-k_1} \ket{c} + \frac{2N_1\sqrt{N_2-k_2}}{F_-} \ket{d} \right) \\
        &\approx \frac{1}{C_-} \left( \sqrt{k_1} \ket{a} + \frac{2N_1\sqrt{k_2}}{F_-} \ket{b} + \sqrt{N_1} \ket{c} + \frac{2N_1\sqrt{N_2}}{F_-} \ket{d} \right) \\
        &= \frac{1}{C_-} \frac{2N_1\sqrt{N_2}}{F_-} \left( \frac{F_-\sqrt{k_1}}{2N_1\sqrt{N_2}} \ket{a} + \frac{\sqrt{k_2}}{\sqrt{N_2}} \ket{b} + \frac{F_-}{2\sqrt{N_1N_2}} \ket{c} + \ket{d} \right) \\
        &\approx \frac{1}{G_-} \left( \frac{F_-}{2\sqrt{N_1N_2}} \ket{c} + \ket{d} \right) \\
        &= \ket{u},
\end{align*}
where in the fifth line, we used $F_- = O(N)$ from \eqref{eq:Fm}. Now, since $N_1 \ne N_2$, $\ket{a}$ and $\ket{b}$ are nondegenerate eigenvectors of $H_\alpha^{(0)}$, and $\ket{u}$ and $\ket{v}$ are also nondegenerate. So that the system evolves from the initial state to marked vertices, there are two critical values of $\gamma$, one that makes $\ket{a}$ degenerate with $\ket{u}$, and the other that makes $\ket{b}$ degenerate with $\ket{u}$. The value of $\gamma$ that makes $\ket{a}$ and $\ket{u}$ degenerate is
\begin{equation}
    \label{eq:gamma-a}
     \gamma_a = \frac{2}{B+(\alpha-1)(N_1-N_2)} = \frac{2}{F_+} = \frac{F_-}{2N_1N_2},
\end{equation}
where the last equality is because $F_+F_- = 4N_1N_2$.

Next, using degenerate perturbation theory, we can lift the degeneracy by including the next-order corrections of $H_\alpha$,
\[
    H_\alpha^{(1)} = -\gamma \begin{pmatrix}
	   0 & 0 & 0 & \sqrt{k_1 N_2} \\
	   0 & 0 & \sqrt{k_2 N_1} & 0 \\
	   0 & \sqrt{k_2 N_1} & 0 & 0\\
	   \sqrt{k_1 N_2} & 0 &  0 & 0 \\
    \end{pmatrix}.
\]
That is, we are solving for linear combinations 
\[ \alpha_a \ket{a} + \alpha_u \ket{u} \]
that satisfy the eigenvalue relation 
\[ \left( H_\alpha^{(0)} + H_\alpha^{(1)} \right)( \alpha_a \ket{a} + \alpha_u \ket{u} ) = E ( \alpha_a \ket{a} + \alpha_u \ket{u} ). \]
In the $\{ \ket{a}, \ket{u} \}$ basis, when $\gamma = \gamma_a = 2/F_+$ \eqref{eq:gamma-a}, this relation is
\[
    \begin{pmatrix}
        -1 + \frac{2N_2(\alpha-1)}{F_+} & -\frac{2\sqrt{k_1N_2}}{F_+G_-} \\
        -\frac{2\sqrt{k_1N_2}}{F_+G_-} & -\frac{4F_-N_1 + (\alpha-1)(4N_1^2+F_-^2)}{2F_+G_-^2N_1} \\
    \end{pmatrix} \begin{pmatrix}
        \alpha_a \\
        \alpha_u \\
    \end{pmatrix} = E \begin{pmatrix}
        \alpha_a \\
        \alpha_u \\
    \end{pmatrix}.
\]
Solving this, we get the following eigenvectors and eigenvalues of $H^{(0)} + H^{(1)}$:
\[
    \begin{pmatrix}
        \alpha_a \\
        \alpha_u
    \end{pmatrix} = \frac{1}{\sqrt{2}} \begin{pmatrix}
        1\\
        \pm 1
    \end{pmatrix}, \quad E = \frac{-B J \mp \sqrt{k_1N_1 B F_- (B F_- - 2 N_1 N_2)}}{B N_1 F_-},
\]
where
\[ J = -F_- (\alpha-1)^2 (N_1 - N_2) + N_1[B + (\alpha-1)(3N_1 - N_1)]. \]
Combining these with the nondegenerate eigenvectors of $H_\alpha^{(0)}$, i.e., $\ket{b}$ and $\ket{v}$, the asymptotic eigenvectors of the search Hamiltonian when $\gamma=\gamma_a$ are
\begin{align*}
    &\ket{\psi_0}=\frac{1}{\sqrt{2}}(\ket{a}+\ket{u)}, \\
    &\ket{\psi_1}=\frac{1}{\sqrt{2}}(\ket{a}-\ket{u)}, \\
    &\ket{\psi_2}=\ket{b}, \\
    &\ket{\psi_3}=\ket{v},
\end{align*}
with corresponding eigenvalues
\begin{align*}
    &E_0 = \frac{-B J - \sqrt{k_1N_1 B F_- (B F_- - 2 N_1 N_2)}}{N_1 B F_-}, \\
    &E_1 = \frac{-B J + \sqrt{k_1N_1 B F_- (B F_- - 2 N_1 N_2)}}{N_1 B F_-}, \\
    &E_2 = -1 - N_1(\alpha-1)\gamma_a, \\
    &E_3 = -\frac{1}{2} [N (\alpha-1) - B] \gamma_a,
\end{align*}
as reported in the main text.


\section{Asymptotic Initial State}

Let us consider how the initial state $\ket{s_{L_\alpha}} = \ket{v_3}$ \eqref{eq:appendix-s} behaves when $\alpha$ is large in magnitude. Taylor expanding around $|\alpha| \to \infty$,
\[ B \approx |(\alpha-1)(N_1 - N_2)| + \frac{2N_1N_2}{|(\alpha-1)(N_1-N_2)|}. \]
Then, for large $|\alpha|$,
\begin{align*}
    F_\pm 
        &\approx |(\alpha-1)(N_1 - N_2)| + \frac{2N_1N_2}{|(\alpha-1)(N_1-N_2)|} \pm (\alpha-1)(N_1-N_2) \\
        &\approx |\alpha(N_1 - N_2)| + \frac{2N_1N_2}{|\alpha(N_1-N_2)|} \pm \alpha(N_1-N_2) \\
        &= \left\{ 1 \pm \text{sgn}[\alpha(N_1-N_2)] \right\} |\alpha(N_1-N_2)| + \frac{2N_1N_2}{|\alpha(N_1-N_2)|},
\end{align*}
and so
\[
    F_+ \approx \begin{cases}
        2\alpha(N_1-N_2), & \alpha(N_1-N_2) > 0, \\
        \frac{2N_1N_2}{|\alpha(N_1-N_2)|}, & \alpha(N_1-N_2) < 0, \\
    \end{cases}
\]
and
\begin{equation}
    \label{eq:Fm}
    F_- \approx \begin{cases}
        \frac{2N_1N_2}{\alpha(N_1-N_2)}, & \alpha(N_1-N_2) > 0, \\
        2|\alpha(N_1-N_2)|, & \alpha(N_1-N_2) < 0. \\
    \end{cases}
\end{equation}
Next, for large $|\alpha|$,
\[
    C_+ \approx \begin{cases}
        \sqrt{N_1} + \frac{N_1^2 N_2}{2\alpha^2\sqrt{N_1}(N_1-N_2)^2}, & \alpha(N_1-N_2) > 0, \\
        \sqrt{N_1 + \frac{\alpha^2(N_1-N_2)^2}{N_2}}, & \alpha(N_1-N_2) < 0, \\
    \end{cases}
\]
and
\[
    C_- \approx \begin{cases}
        \sqrt{N_1 + \frac{\alpha^2(N_1-N_2)^2}{N_2}}, & \alpha(N_1-N_2) > 0, \\
        \sqrt{N_1} + \frac{N_1^2 N_2}{2\alpha^2\sqrt{N_1}(N_1-N_2)^2}, & \alpha(N_1-N_2) < 0. \\
    \end{cases}
\]

Now, we want to show that the initial state $\ket{s_{L_\alpha}} = \ket{v_3}$ \eqref{eq:appendix-s} asymptotically approaches a uniform superposition over the left vertices, or right vertices, depending on the sign of $\alpha(N_1-N_2)$. That is, we want to examine the amplitudes $1/C_-$ and $2N_1/(C_-F_-)$.

When $\alpha(N_1-N_2) \to -\infty$,  such as when optimally searching for a marked vertex in the left partite set with $\gamma = \gamma_a$,
\[ \frac{1}{C_-} \to 1/\sqrt{N_1}, \]
and
\[ \frac{2N_1}{C_-F_-} \approx \frac{2N_1}{\sqrt{N_1} 2|\alpha(N_1-N_2)|} \to 0, \]
and so the initial state \eqref{eq:appendix-s} becomes
\[ \frac{1}{\sqrt{N_1}} \sum_{i\text{ left}} \ket{i}, \]
which is a uniform superposition over the vertices in the left partite set.

When $\alpha(N_1-N_2) \to \infty$, such as when optimally searching for a marked vertex in the right partite set with $\gamma = \gamma_b$,
\[ \frac{1}{C_-} \to 0, \]
and
\[
    \frac{2N_1}{C_-F_-}
        \approx \frac{2N_1}{\sqrt{N_1 + \frac{\alpha^2(N_1-N_2)^2}{N_2}} \frac{2N_1N_2}{\alpha(N_1-N_2)}} 
        \approx \frac{1}{N_2\sqrt{\frac{N_1}{\alpha^2(N_1-N_2)^2} + \frac{1}{N_2}}}
        \to \frac{1}{\sqrt{N_2}},
\]
and so the initial state \eqref{eq:appendix-s} becomes
\[ \frac{1}{\sqrt{N_2}} \sum_{i\text{ right}} \ket{i}, \]
which is a uniform superposition over the vertices in the right partite set.


\section{Derivative of the Runtime}

From the main text, the runtime is
\[ t_*=\frac{\pi}{\Delta E} = \frac{\pi}{2} \sqrt{\frac{N_1 B F_-}{k_1 (BF_--2N_1N_2)}}. \]
Differentiating with respect to $\alpha$ using the chain rule and quotient rule, we get
\begin{align*}
    \frac{dt_*}{d\alpha} 
        &= \frac{\pi}{4} \left( \frac{N_1 B F_-}{k_1 (BF_--2N_1N_2)} \right)^{-1/2} \frac{1}{k_1} \frac{(BF_- - 2N_1N_2)\frac{d}{d\alpha}(N_1BF_-) - N_1BF_-\frac{d}{d\alpha}(BF_- - 2N_1N_2)}{(BF_- - 2N_1N_2)^2} \\
        &= \frac{\pi}{4} \sqrt{\frac{k_1 (BF_--2N_1N_2)}{N_1 B F_-}} \frac{1}{k_1} \frac{(BF_- - 2N_1N_2) N_1 \frac{d}{d\alpha}(BF_-) - N_1BF_- \frac{d}{d\alpha}(BF_-)}{(BF_- - 2N_1N_2)^2} \\
        &= \frac{\pi}{4} \sqrt{\frac{k_1 (BF_--2N_1N_2)}{N_1 B F_-}} \frac{1}{k_1} \frac{- 2N_1^2N_2 \frac{d}{d\alpha}(BF_-)}{(BF_- - 2N_1N_2)^2} \\
        &= -\frac{\pi}{2} \frac{N_2}{\sqrt{k_1 B F_-}} \left( \frac{N_1}{BF_- - 2N_1N_2} \right)^{3/2} \frac{d}{d\alpha}(BF_-).
\end{align*}
So, we need the derivative of $BF_-$. Using the product rule,
\begin{align*}
    \frac{d}{d\alpha}(BF_-)
        &= \frac{dB}{d\alpha} F_- + B \frac{dF_-}{d\alpha} \\
        &= \frac{dB}{d\alpha} F_- + B \frac{d}{d\alpha} \left[ B - (\alpha-1)(N_1-N_2) \right] \\
        &= \frac{dB}{d\alpha} F_- + B \left[ \frac{dB}{d\alpha} - (N_1-N_2) \right] \\
        &= \frac{dB}{d\alpha} \left( B + F_- \right) - B (N_1-N_2).
\end{align*}
Now, the derivative of $B$ is
\begin{align*}
    \frac{dB}{d\alpha} 
        &= \frac{d}{d\alpha} \sqrt{(\alpha-1)^2(N_1-N_2)^2+4N_1N_2} \\
        &= \frac{(\alpha-1)(N_1-N_2)^2}{\sqrt{(\alpha-1)^2(N_1-N_2)^2+4N_1N_2}} \\
        &= \frac{(\alpha-1)(N_1-N_2)^2}{B},
\end{align*}
and plugging this into the derivative of $BF_-$, we get
\begin{align*}
    \frac{d}{d\alpha}(BF_-)
        &= \frac{(\alpha-1)(N_1-N_2)^2}{B} \left( B + F_- \right) - B (N_1-N_2) \\
        &=  (N_1-N_2) \frac{(\alpha-1)(N_1-N_2)\left( B + F_- \right) - B^2}{B} \\
        &=  (N_1-N_2) \frac{(\alpha-1)(N_1-N_2)\left[ 2B - (\alpha-1)(N_1-N_2) \right] - B^2}{B} \\
        &=  (N_1-N_2) \frac{2B(\alpha-1)(N_1-N_2) - (\alpha-1)^2(N_1-N_2)^2 - B^2}{B} \\
        &=  -\frac{(N_1-N_2)F_-^2}{B}.
\end{align*}
Plugging this into the derivative of the runtime,
\begin{align*}
    \frac{dt_*}{d\alpha}
        &= -\frac{\pi}{2} \frac{N_2}{\sqrt{k_1 B F_-}} \left( \frac{N_1}{BF_- - 2N_1N_2} \right)^{3/2} \frac{-(N_1-N_2)F_-^2}{B} \\
        &= \frac{\pi}{2} \frac{(N_1-N_2)N_2}{\sqrt{k_1}} \left[ \frac{N_1F_-}{B(BF_- - 2N_1N_2)} \right]^{3/2}.
\end{align*}
We can clean this up further by simplifying $B F_- - 2N_1N_2$, which from the definition of $F_-$ is
\begin{align*}
    B F_--2N_1N_1
        &= B[B - (N_2-N_1)(\alpha-1)] - 2N_1N_2 \\
        &= B^2+ - (N_2-N_1)(\alpha-1) -2N_1N_2.
\end{align*}
We can express the last term, $2N_1N_2$, in terms of $B^2$. From the definition of $B$,
\[ B^2 = (N_2-N_1)^2(\alpha-1)^2+4N1N_2, \]
or
\[ 4N_1N_2 = B^2-(N_2-N_1)^2(\alpha-1)^2, \]
so
\[ 2N_1N_2 = \frac{1}{2}[B^2-(N_2-N_1)^2(\alpha-1)^2]. \]
Then we can substitute this back and get 
\begin{align*}
    B F_--2N_1N_1
        & = B^2 - B(N_2-N_1)(\alpha-1) - \frac{1}{2}[B^2-(N_2-N_1)^2(\alpha-1)^2] \\
        & = \frac{B^2}{2} - B(N_2-N_1)(\alpha-1) + \frac{1}{2}(N_2-N_1)^2(\alpha-1)^2 \\
        & = \frac{1}{2}[B^2 - 2B(N_2-N_1)(\alpha-1) + (N_2-N_1)^2(\alpha-1)^2] \\
        &= \frac{1}{2}[B - (N_2-N_1(\alpha-1)]^2 \\
        &= \frac{F_-^2}{2}
\end{align*}
Plugging this into our expression for $dt_*/d\alpha$ from earlier, we get      
\begin{align*}
    \frac{dt_*}{d\alpha}
        &= \frac{\pi}{2} \frac{(N_1-N_2)N_2}{\sqrt{k_1}} \left( \frac{N_1F_-}{BF_-^2/2} \right)^{3/2} \\
        &= \frac{\pi \sqrt{2} N_1^{3/2}N_2(N_1-N_2)}{\sqrt{k_1}B^{3/2}F_-^{3/2}},
\end{align*}
as reported in the main paper.